\documentclass[preprint,superscriptaddress,showpacs,aps,prl]{revtex4}
\usepackage{makeidx}
\usepackage{bm}
\usepackage{graphics}
\usepackage[dvips]{epsfig}
\newcounter{fig}

\begin{document}
\title{ Replay to  "Comment on 'Screening in gated bilayer graphene' "}
\author{L.A. Falkovsky}
\affiliation{L.D. Landau Institute for Theoretical Physics, Moscow
119334, Russia}
\pacs{73.20.At, 73.21.Ac, 73.43.-f}

\date{\today}      

\begin{abstract} We discuss the physics of the tunable bandgap in
bilayer graphene with the gate voltage and doping. A comparison
with  experimental data obtained by Kuzmenko et al [Phys. Rev. B
\textbf{80}, 165406 (2009)] demonstrates the good agreement.

\end{abstract}
\maketitle

Bilayer graphene is widely known (see, for instance, \cite{OP}) as a tunable bandgap
semiconductor: its band structure depends on the external gate
voltage and doping. This phenomenon is promising for application.
From the theoretical point of view, the problem can be considered
as follows.  Two parameters, the chemical potential and the bandgap,
should be determined in an external electric field, normal to the bilayer
surface. Then, we have a typical variation problem.
As the effect of doping was not considered in the previous
publication \cite{Mc}, we solved this problem \cite{Fal} with
the method of the total energy minimization. The problem was considered as well in Ref. \cite{GLS} within DFT calculations.

The result \cite{Fal} does not coincide with Ref. \cite{Mc} even
for the case of undoped bilayer graphene. The reason of the
disagreement is in the definition of the ground state of the
system. In Ref. \cite{Fal}, we assume that the ground state is realized in the undoped
pristine bilayer where the chemical potential is situated between two nearest bands.
Therefore, at doping or gate
voltage, we take into account only the excitation  of  holes
in the valence band or electrons in the conduction band. Of course, this concept takes implicitly into account
electron-electron interactions. In contrast to this, while considering the effect of the external electric field  in Refs. \cite{Mc, Mc1},
the energy of excitations in the  completely filled deep
band is included in the total energy. First of all, such a consideration,   involving the large
contribution of completely filled band, has no need  in the variation method in contradictions with the statement of the paper \cite{Mc1}. The method used in \cite{Mc1} gives only 
the evident electrostatic condition, Eq. (13). This condition is naturally applied from the outset
in Ref. \cite{Fal} to construct the total energy.  Second, the including the energy of the deep
states violates the concepts of the normal Fermi liquid,
according to which  the responses of the Fermi liquid are resulted from the neighborhood of  the Fermi surface. 

Several statements of Ref. \cite{Mc1} are incorrect, including for example,
the attempt  [see, paragraph after Eq. (13)] to extract  the direction
of the electric field from the scheme in Fig. 3 of Ref.
\cite{Fal}. The authors of Ref. \cite{Mc1} have forgotten that the electric field is determined by carriers as well as dopant which does not shown in the scheme.  In fact, two possible directions of the electric
field are consistent with two signs in Eq. (14) of Ref. \cite{Fal}
and the solution to this equation is found only at the certain sign
that chooses  the field direction.

The results of the Ref. \cite{Mc1} differ from the previous paper \cite{Mc} only in the involving of dopant. The authors statement before Eq. (2) is  wrong, no such formula was derived in their paper \cite{Mc}.
The authors of the paper \cite{Mc1} do not
present any comparison of their results with experiments
\cite{KCM,ZTG,MLS} referring the  disorder as a reason  of possible disagreement of their theory with experiments.
For definiteness, we compare  in Fig. 1 the result of Ref. \cite{Fal} with
experimental data from Ref. \cite{KCM}. The complicated behavior near low carrier concentrations is connected with the effect of doping. Other result of the doping, the asymmetry at the electron-hole sides, is evident from the figure.  

\begin{figure}[]
\resizebox{.5\textwidth}{!}{\includegraphics{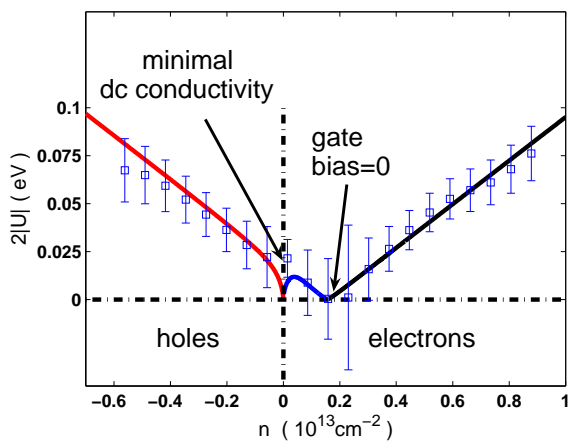}}
\caption{The gap in  eV versus the carrier concentration for the
electron doping with the concentration $N_2=0.78\times10^{12}$
cm$^{-2}$ (our theory); the positive (negative) values of $n$
correspond to the electron (hole) conductivity; squares are
experimental data \cite{KCM}.} \label{ukuzexp.eps}
\end{figure}

This work was supported by the Russian Foundation for Basic
Research (grant No. 10-02-00193-a).

\end{document}